\documentclass[12pt]{iopart}
\usepackage{iopams}
\usepackage{graphicx,graphics} 
\usepackage{supertabular}
\usepackage{times}
\usepackage{bm} 
\usepackage{subfigure} 
\usepackage{ulem}

\begin{document}




\title[Quantum Imager]{A Quantum Imager for Intensity Correlated Photons}

\author{D L Boiko,$^{1,2}$ N J Gunther,$^3$ N Brauer,$^1$ M Sergio,$^1$ C Niclass,$^1$ G B Beretta,$^4$ and E Charbon$^1$}
\address{
$^1$Ecole Polytechnique F\'ed\'erale de Lausanne, 1015, Lausanne, Switzerland \\
$^2$Centre Suisse d'Electronique et de Microtechnique SA, 2002,
Neuch\^atel, Switzerland \\
$^3$Performance Dynamics, 4061 East Castro Valley Blvd., Suite
110, Castro Valley, California, USA \\
$^4$HP Laboratories, 1501 Page Mill Road, Palo Alto, California,
USA} \ead{dmitri.boiko@csem.ch}






%


\begin{abstract}
We report on a 
device capable of imaging second-order spatio-temporal
correlations $g^{(2)}(\mathbf x, \tau)$ between photons. The
imager is based on a monolithic array of single-photon avalanche
diodes (SPADs) implemented in CMOS technology and a simple
algorithm to treat multiphoton time-of-arrival distributions from
different SPAD pairs. It is capable of $80$ ps temporal
resolution with fluxes as low as $10$ photons/s at room
temperature. An important application might be the local imaging
of $g^{(2)}$ as a means of confirming the presence of true
Bose-Einstein macroscopic coherence (BEC) of cavity exciton
polaritons.
\end{abstract}

\submitto{\NJP}

\maketitle

Recent
experiments~\cite{Kasprzak06,Deng06,Christopoulos07,Balili07}
have reported the Bose-Einstein condensation (BEC) phase
transition in polariton system in a semiconductor microcavity. The
macroscopic quantum degeneracy is typically detected by probing
the statistical properties of light emitted from a microcavity,
under the presumption that the statistics of the exciton
polaritons are faithfully transferred to the emanating photons.
\begin {figure}[tbf]
\begin{center}
\includegraphics [scale = 1.35] {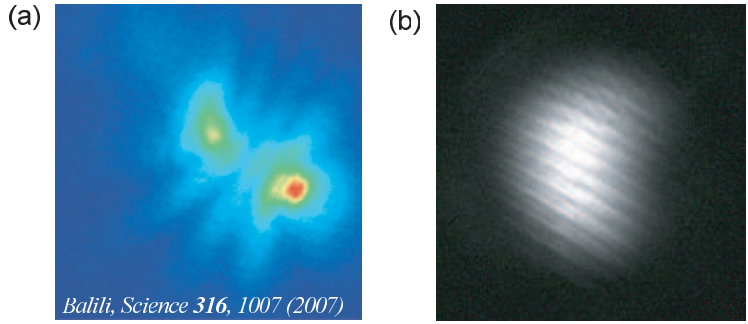} 
\caption { Interference fringes (a) at $\lambda=770$ nm
wavelength used to verify the BEC of polaritons in GaAs
microcavity (reproduced from Ref.~\cite{Balili07}) and (b) at
$\lambda=546$ nm measured for the green line of pulsed Hg-Ar
discharge lamp.} \label {FIG0}
\end{center}
\end {figure}

It has been further assumed that observation of the
interference fringes similar to those in Michelson or Young
interferometers (figure \ref{FIG0}(a)) is sufficient to establish
the fact of macroscopic coherence in polariton
system.~\cite{Kasprzak06,Balili07,Snoke03} Two points on the wave
front separated by a distance $x_{12}$  produce intensity pattern
$I_1+I_2+2 \sqrt{I_1I_2} | g^{(1)}(\mathbf{x}_{12},\tau)|
\cos(\Delta \varphi_{12})$, such that the fringe visibility
measures the magnitude of the first-order correlation function.
But simply measuring this quantity alone is ambiguous because a
coherent light source (\textit{e.g.}, a photon laser or decaying
polariton BEC) can exhibit the same first-order correlations as a
chaotic (or thermal) light source (\textit{e.g.} Hg-Ar discharge
lamp in figure \ref{FIG0}(b)). Table~\ref{PhotStates} shows that
proper disambiguation of a coherent state also requires
measurement of the second-order correlation function
%
$g^{(2)}(\mathbf{x}_{12}, \tau) {=} \frac{\langle I_1(t) \, I_2(
t+\tau)\rangle}{\langle I_1(t)\rangle \langle I_2(t) \rangle }$
associated with intensity noise correlations. Here, $I_{1,2}(t)$
is the light intensity at a point $\pm \frac12 \mathbf{x}_{12}$
and time $t$. The minimal condition to confirm the BEC phase
transition in a polariton system then becomes
\mbox{$g^{(1)}(\mathbf x, 0)=g^{(2)}(\mathbf x,0)=1$} (third
column of Table~\ref{PhotStates}). So far, very few theoretical
and experimental studies of the second-order correlations in a
polariton system were limited to the $k{=}0$ point in the
momentum space (in the lateral cavity direction), reporting thus
a spatially-averaged value of $g^{(2)}(0)$ and ignoring the fact
that 2D BEC can be achieved only in spatially confined systems.
As a consequence, the model of Ref.\cite{Schwendimann08} predicts
increasing correlation peak height $g^{(2)}(0)$ with polariton
number caused by the strong scattering effects above the critical
threshold density. Such behaviour for polaritons at $k=0$ has
been confirmed in Ref.\cite{Kasparzak08}, while the experimental
results of Ref.\cite{Deng02} disagree with such behaviour. To
confirm the BEC phase transition in a polariton system, one needs
to distinguish the presence of both coherent condensate
($g^{(1)}{=}g^{(2)}$=1) and thermal noncondensate ($g^{(1)}
{\neq} g^{(2)}$) fractions, which can be achieved by local
measurement of the first-order and second-order correlations.
However, the small dimensions of the polariton distribution in
the microcavity, means that such measurements must be capable of
resolving a spatial dependence in $g^{(2)}$. All these
requirements demand an integrated monolithic photon detector,
like a camera, but capable of imaging intensity noise
correlations.


\Table{\label{PhotStates} Values of first and second order
correlation functions for incoherent, coherent and thermal light
states.}
\br
Function & Incoherent  & Coherent & Chaotic  \\
\mr
$g^{(1)}(\mathbf x,0)$ & 0 & 1 & 1 \\
$g^{(2)}(\mathbf x,0)$ & 1 & 1 & 2 \\
\br
\endTable

\begin{figure}[tbf]
\begin{center}
\includegraphics[scale = 1.4]{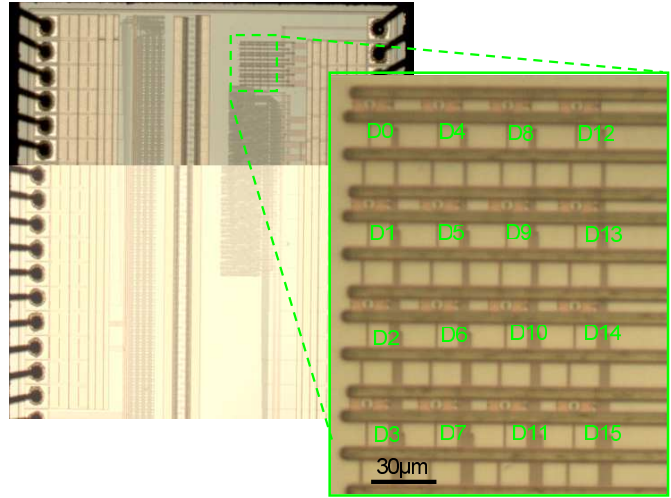}
\caption{ Exploded micrograph of the $4\times4$ SPAD array.}
\label{fig:fig1}
\end{center}
\end{figure}

In this letter, we present an imaging device allowing the
second-order (intensity) correlations of a field to be probed
locally. The current implementation 
comprises a
$4\times4$ silicon array of single-photon avalanche diodes
(SPADs) implemented in 0.35 $\mu m$ CMOS technology ~\cite{Niclass06}. The device incorporates on-chip high
bandwidth I/O circuitry (figure \ref{fig:fig1}) for off-chip data
processing of multi photon arrivals. We demonstrate the
operational performance of our monolithic $g^{(2)}$-imager in a
miniature replication of the Hanbury Brown and Twiss (HBT)
stellar interferometer ~\cite{HBT56A}.

Each pixel in our SPAD array is based on an avalanche photodiode
operating in the Geiger mode.
The $3.5\mu m$-diameter active region of a
SPAD n-well pixel consists of a p$^{+}$-n junction reverse-biased above its breakdown voltage.
When a photon is absorbed in the multiplication region, an
avalanche is triggered with a certain probability.
The measured single photon detection probability of such 0.35$\mu m$-CMOS SPAD pixels is 25 $\%$
at 546 nm wavelength and 4 V excess voltage above the breakdown threshold.
(The detection probability is 40\%  at the peak sensitivity wavelength 450 nm ~\cite{Niclass06}.) The avalanche breakdown is
subsequently quenched by an on-chip ballast resistor, which is used to read out the photodetection
events. Its value defines the dead time of the detector which is 15 ns for each SPAD pixel of the 4x4 array considered here.
The chip comprises built-in
high-bandwidth electronics to convert Geiger pulses into digital
signals for off-chip data processing. This design drastically
improves the signal-to-noise ratio. At room temperature, the
lowest detectable photon flux is set by the dark count rate (DCR)
of SPADs in the $5$--$10$ Hz range. Such low DCR is achieved by
using small n-wells of diameter $3.5~\mu m$. The lowest
detectable photon flux density in our experiments is
${\sim}10^8 photons/s {\cdot} cm^2$, well below the level in polariton
BEC experiments. The array pitches are $30~\mu m$ horizontal and
$43~\mu m$ vertical. All $16$ detectors in the array have
separate parallel outputs so that 
$ \biggl( \begin{array}{c} 16 \\ 2 \end{array} \biggr) = 120$
simultaneous pairwise measurements are possible at a temporal
resolution limited by the SPAD jitter characteristics ($80$ ps).

%
%

\begin {figure}[tbf]
\begin{center}
\includegraphics [scale = 1.4]{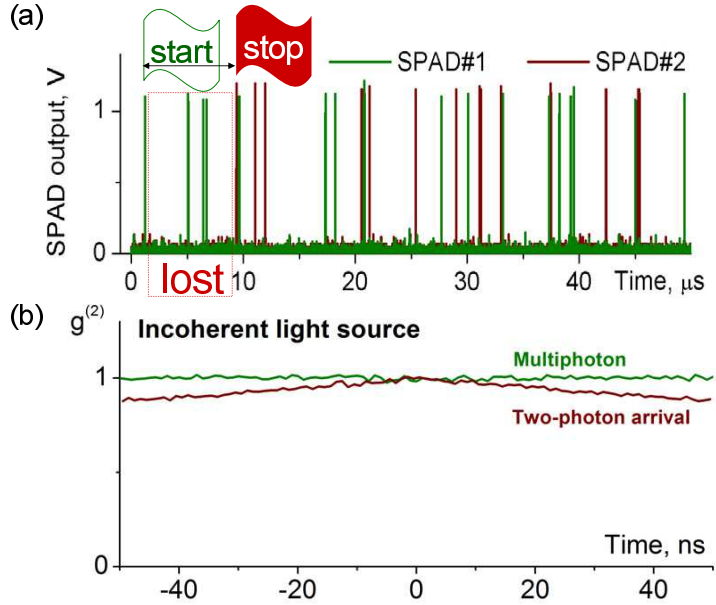} 
\caption {Comparison of single-photon and multi-photon
measurements. (a) Traces from two detectors at 5 MHz count rate.
Standard histogram of delayed single-photon arrivals shows a
large number of lost detections. (b) Computed $g^{(2)}$ based on
multi-photon arrivals (green) and the standard technique (red)
showing the impact of lost detections at count rates $\mu \sim$2.5 MHz at each detector and the width of measurement window $NT$=100ns.} \label{fig2}
\end{center}
\end{figure}

The pairwise intensity noise correlations $g^{(2)}(\mathbf
x_{ij},\tau)$ are computed using an external four-channel $6$ GHz
bandwidth digital oscilloscope (Wavemaster 8600A, LeCroy) by
programming it with an algorithmic version of equation
(\ref{g2Num}):
\begin{equation}
g^{(2)}
(\mathbf {x}_{ij}{,} \tau) 
{=}\frac{ {\displaystyle
NM{\sum_{m=0}^M} 
\sum_{n{=}-\frac N 2}^{\frac N 2} X_i^{(m)}(n) 
X_j^{(m)}(n{+}l) } } { {\displaystyle {\sum_{m{=}0}^M}
\sum_{n{=}{-} \frac N 2}^{\frac N 2} X_i^{(m)}(n)
{\sum_{n'{=}{-}\frac N 2}^{\frac N 2}} X_j^{(m)}(n'{+}l) } }
\label{g2Num}
\end{equation}
where integers $i,j  \;(i \neq j)$ enumerate detector pixels,
$X_i$ and $X_j$ are discrete random variables whose values $0$
(no event) or $1$ (photon detection) correspond to the binary
data stream (figure \ref{fig2}(a)) emanating from any pair of
detectors $D_i$ and $D_j$, respectively. The spatial lag $x_{ij}$
is set by the separation of the detector pair within the SPAD
array. Time-lag increments $\tau=lT$  are set by multiples of
temporal resolution $T$, where $NT$ is the width of measurement
window and $M$ is the overall number of measurements series.

\begin{figure}[tbf]
\begin{center}
\includegraphics [scale = 1.4] {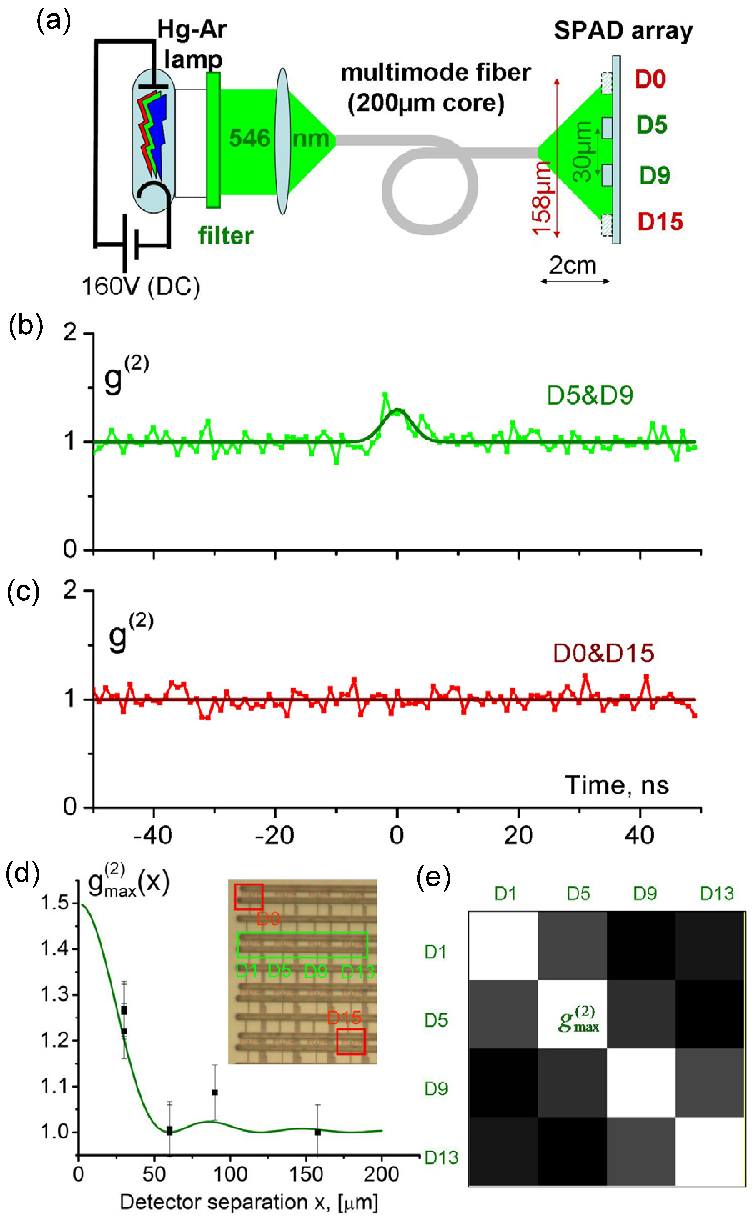} 
\caption {(a) SPAD array configured as miniature HBT
interferometer for measuring $200~\mu$m (b) correlations between
adjacent detectors ($x_{5,6}=30\mu m$) and (c) extreme diagonal
detectors ($x_{0,15}=158\mu m$). (d) Measured (points) and
calculated (curve) second-order in function of detector
separation The micrograph inset shows the position of detectors
within the array. (e) imaged second-order correlation maxima
along the row of array in (d). It is assumed that $g^{(2)}=2$
along the diagonal.} \label{fig4}
\end{center}
\end {figure}

Unlike conventional detection methods based on start-stop timing
histograms of delayed single photon arrivals [figure
\ref{fig2}(a)], our approach implements properly normalized
multiphoton distribution which is robust against missing
detection events, the impact of Poisson-like distribution decay
${\propto} \exp ({-}\mu \tau) $ at large $\tau$ and intensity
modulation. Equation (\ref{g2Num}) permits any count rates and
temporal window of interest and does not require a statistical
hypothesis to normalize $g^{(2)}$. Figure \ref{fig2}(b) shows the
benefits of our technique by comparing $g^{(2)}(\tau)$
measurements of incoherent broad-band light using multiphoton
arrivals and the standard two-photon histogram.

In figure \ref{fig2}(b), for the standard approach based on
time-delayed coincidence events, the histogram was
acquired and then normalized assuming that at the lag $\tau{=}0$
there should be no correlations for incoherent light source
(\textit{i.e.}, assuming that $g^{(2)}(0){=}1$).
We shall remind that a probability distribution of time intervals between two consecutive 
photons
is $p(0|\tau)=\mu \exp(-\mu \tau)$ for a Poisson process \cite{Glauber07, Saleh73} and that for a multi-mode Gaussian (chaotic light) and a coherent (Poisson statistics) states these distributions 
are the same  \cite{Arecchi65}.
Therefore, when normalized
in such way, the standard method shows the onset of Poisson
distribution $g^{(2)}(\tau)=p(0|\tau)/\mu \simeq 1{-} \mu |\tau|$ for large time
lag $|\tau| \sim 0.1\mu^{-1}$. Thus in fig \ref{fig2}(b), the average count rate at detectors is $\mu=2.5$ MHz, and for $\tau =50$ ns, the error is $\Delta g^{(2)} \sim 0.13 $.
The
conventional procedure is thus limited by the small count rate
and temporal window width product. At the same time, our approach
produces the correct correlations independent of the measurement
interval $NT$ and the photon flux intensity.

The $g^{(2)}$-imager was tested by measuring the statistical
properties of an extended thermal light source. As a model system
for a quasi-monochromatic chaotic light source, we used a Hg-Ar
spectral wavelength calibration lamp (bulb CAL-2000-B, Ocean
Optics) with U-folded discharge and cold cathode. This lamp,
designed for operation in the AC regime, also operated well with
a DC power supply ($160$ V @ $15$ mA). To start the discharge, we
used the original AC power supply of CAL2000 source which was
connected in parallel with a DC source via a filter ($2 H$
inductance) such that the AC supply was gradually turned off
while the DC source was gradually turned on.

The light emitted by the lamp was transmitted through a $10$ nm
bandpass filter (FL543.5-10, Thorlabs), which keeps only the
emission at the green line of mercury ($546$ nm). The light then
was injected into a $1$ m long multimode fiber (figure
\ref{fig4}(a)) with core diameter $w{=}200~\mu m$.  The other end
of the fiber was used to illuminate the SPAD array placed in the
far field zone of the fiber end, at a distance $L{=}2$ cm from
the fiber [figure \ref{fig4}(a)]. The numerical aperture of the
fiber (NA=0.22) assumes that the whole $4 \times 4$ SPAD array is
over illuminated. Such extended thermal light source is of the
angular width $w/L{=}10^{-2}$ rad and exhibits first-order
correlations $g^{(1)}$ (figure \ref{FIG0}(b)), when SPAD array is
replaced by Young double-pinhole interferometer (not shown in the
figure).

The second-order spatio-temporal correlations for a non-polarized
single-mode chaotic 
light source are determined by the first-order
correlations~\cite{Glauber07,Glauber63a,Glauber63b} with coherence
time $\tau_c{=}2\sqrt{2 \pi \ln2}/\Delta \omega$ due to the
inhomogeneous broadening $\Delta \omega$ (FWHM) of the emission
line
\begin{equation}
g^{(2)}(\bm x_{ij}, \tau) =1+\frac 12 \left|g^{(1)}(\bm
x_{ij},\tau)\right|^2  \\
=1+\frac 12{\rm sinc}^2 \Bigl(\frac{\pi w}{\lambda L} x_{ij}
\Bigr) \, \exp \Bigl(-\pi \frac{\tau^2}{\tau_c^2} \Bigr) \, ,
\label{g1full}
\end{equation}
where the second term in the right hand side takes into account
the decorrelation effects due to unpolarized light (the
coefficient $1/2$), the zero-delay degree of spatial coherence and
the Gaussian profile of the delayed first-order correlation
function for an inhomogeneously-broadened line.

To examine the temporal correlations, we used detector pairs in
the middle of the array as well pairs at diagonally opposite
corners, thus providing correlations $g^{(2)}(x_{ij}, \tau)$
between two different regions separated by $30$ and $158~\mu m$,
respectively. At small separation (figure \ref{fig4}(b)),  the
data shows an excess of coincidences $g^{(2)}(x_{5,9}, 0) {=}
1.25$. At large distance, detector counts are uncorrelated
(figure \ref{fig4}(c)).

At temporal resolution of the scope $T{=}1$ ns, the measured
coherence time $\tau_c$ is 5.2 ns such that the impact of
integration effects on measured correlation peak width and height
is less than 2 $\%$. \cite{Scarl66} The corresponding linewidth
of the source is $130~MHz$ (FWHM), which is less than the Doppler
width of the green line of Hg and can be attributed to the Dicke
linewidth narrowing due to the buffering effect of Ar in the lamp
bulb.

The spatial oscillations of $g^{(2)}(x_{ij}, 0)$ due to the
spatial coherence factor in (\ref{g1full}) were measured by
selecting detector pairs from a row of the array (D1--D13 row in
figure \ref{fig:fig1}) with detector separation 30, 60 and 90
$\mu$m. Figure \ref{fig4}(d) shows the correlation excess at
zero-delay is well fitted by a $\rm sinc$ function in equation
(\ref{g1full}), yielding the angular width of the source
$0.9{\times} 10^{-2}$ rad, close to the estimated value.

Being limited by the number of acquisition channels, we were able
to record simultaneous correlations between four independent
detectors. In figure \ref{fig4} (e), $g^{(2)}(\mathbf x_{ij}, 0 )$
measured along the array row is plotted as a pairwise correlation
map $g^{(2)}(i,j)$. In this image map, the spatial oscillations
of the coherence factor are clearly visible.

In conclusion, we have presented a 
$g^{(2)}$-imager built with conventional CMOS technology, which
is capable of measuring second-order spatio-temporal correlated
photons and thereby offers an important means for verifying the
existence of a BEC state of cavity exciton polaritons. Future
work will include the development of larger arrays of SPADs, the
integration of on-chip data processing based on equation
(\ref{g2Num}), and the extension to other $g^{(2)}$-imaging
applications.

\ack
 This research was supported, in part, by a grant of the Swiss
National Science Foundation.




%
%
%
%
%
%
%
%
%

\end{document}